\documentclass[ amsmath,amssymb,aps,prb,reprint,superscriptaddress,showpacs]{revtex4-1}

\usepackage{graphicx}% Include figure files
\usepackage{dcolumn}% Align table columns on decimal point
\usepackage{bm}% bold math
\usepackage{amsmath}

\begin{document}

%Title of paper
\title{Highly mobile carriers in orthorhombic phases of iron-based superconductors  FeSe${}_{1-x}$S${}_{x}$.}

\author{Y.A. Ovchenkov}
\email[]{ovtchenkov@mail.ru}
\affiliation{Faculty of Physics, M.V. Lomonosov Moscow State University, Moscow 119991, Russia}
\author{D.A. Chareev}
\affiliation{Institute of Experimental Mineralogy, RAS, Chernogolovka, 123456, Russia}
\affiliation{Ural Federal University, 620002 Ekaterinburg, Russia}
\author{V.A. Kulbachinskii}
\author{V.G. Kytin}
\affiliation{Faculty of Physics, M.V. Lomonosov Moscow State University, Moscow 119991, Russia}
\author{D.E. Presnov}
\affiliation{Skobeltsyn Institute of Nuclear Physics, Moscow 119991, Russia}
\author{O.S. Volkova}
\author{A.N. Vasiliev}
\affiliation{Faculty of Physics, M.V. Lomonosov Moscow State University, Moscow 119991, Russia}
\affiliation{Ural Federal University, 620002 Ekaterinburg, Russia}
\affiliation{National University of Science and Technology 'MISiS' , Moscow 119049, Russia}

\date{\today}
%%%%%%%%%%%%%%%%%%%%%%%%%%%%%%%%%%%%%%%
%
\begin{abstract}
The field and temperature dependencies of the longitudinal and Hall resistivity have been measured for FeSe${}_{1-x}$S${}_{x}$ (x=0.04, 0.09 and 0.19) single crystals.  The sample  FeSe${}_{0.81}$S${}_{0.19}$ does not show a transition to an orthorhombic phase and exhibits at low temperatures the transport properties quite different from those of orthorhombic samples. The behavior of FeSe${}_{0.81}$S${}_{0.19}$ is well described by the simple two band model with comparable values of hole and electron mobility. In particular,  at low temperatures the transverse resistance shows a linear field dependence, the magnetoresistance follow a quadratic field dependence and obeys to Kohler's rule. In contrast, Kohler's rule is strongly violated for samples having an orthorhombic low temperature structure. However, the transport properties of the orthorhombic samples can be satisfactory described by the three band model with the pair of almost equivalent to the tetragonal sample hole and electron bands, supplemented with the highly mobile electron band which has two order smaller carrier number. Therefore, the peculiarity of the low temperature transport properties of the orthorhombic Fe(SeS) samples, as probably of many other orthorhombic iron superconductors, is due to the presence of a small number of highly mobile carriers which originate from the local regions of the Fermi surface, presumably, nearby the Van Hove singularity points.  
\end{abstract}
%
% insert suggested PACS numbers in braces on next line
% insert suggested keywords - APS authors don't need to do this
%
%\maketitle must follow title, authors, abstract, \pacs, and \keywords
\maketitle
%
%
%
%%%%%%%%%%%%%%%%%%%%%%%%%%%%%%%%%%%%%%%%%%%%
\section{Introduction}
A complex behavior of the normal state transport properties in the copper high temperature superconductors (HTSC) have been studied for a long time. Nevertheless, up to now there is no consistent description for some basic transport properties of copper HTSC as, for example, for an unusual temperature dependence of the Hall and longitudinal resistivity. The study of the transport properties of copper HTSC is often complicated by a phase separation and other microstructure peculiarity of studied crystals. This has led to a significant evolution in the views of microscopic electronic properties through the continuous improvements in investigation methods and sample quality. (see for example recent review \cite{1_nature14165} and references therein). In particular, it is strongly influenced by the development of the angle resolved photoemission spectroscopy (ARPES) techniques \cite{2_RevModPhys.75.473} which brought invaluable information about the existence of some specific features of the Fermi surface (FS) structure. Now it is clear that some of the anomalous normal state properties can be due to local kinetics of excitations at some regions of FS \cite{3_PhysRevB.88.041104}.
    
The members of a new family of HTSC, the iron superconductors \cite{4_JACS063355c}, show sometimes very similar anomalous behavior of the normal state transport properties. First of all, the Hall constants, similar to the Hall constants of the copper HTSC, demonstrate a large temperature variation, often with a sharp change in slope below 100 -- 150 K, as was observed for LiFeAs \cite{5_PhysRevB.84.064512}, Ba(FeMe)${}_{2}$As${}_{2}$ (Me=Co,Cu) \cite{6_PhysRevB.80.054517} and other superconducting series. Besides, the Hall resistance and longitudinal magnetoresistance may significantly deviate, correspondingly, from linear and quadratic in magnetic field dependencies as was observed for BaFe${}_{2}$As${}_{2}$ \cite{7_PhysRevB.84.184514}.

In general, the iron superconductors are multiband semimetals. The FS is formed by iron $d$-electrons \cite{8_Ann.Phys.583.8}. The resistivity of compounds is relatively high ( 0.1 -- 10 m${}\Omega$ cm ) and probably can be explained by peculiarities of $d$ electrons in these compounds \cite{9_PhysRevB.87.024504}. The recent ARPES studies of the iron superconductors revealed importance of electronic correlations in achieving a high $T_{c}$'s \cite{PhysRevX.4.031041, PhysRevLett.115.256403}. The observed pronounced reduction of the overall bandwidth on going to superconductivity, band-selective impurity scattering \cite{PhysRevX.4.031041} and other common phenomenon of iron superconductors need further understanding.
 
Similar to copper HTSCs and many other superconducting families the sample microstructure or inhomogeneity often hinders the study of the transport properties of iron superconductors. For example, superconducting (Na,K)Fe${}_{2}$Se${}_{2}$ crystals in normal state show phase separation a metal and semiconducting phases which leads to a formation of domains with a large variation of a local conductivity \cite{10_J.A.P116.043904}. Fortunately, these complexities are minimized for high quality superconducting crystals of FeSe and substituted compositions \cite{11_PhysRevB.88.174512, 12_PNAS105.14262} which show superconductivity in a stoichiometric form.

We have studied temperature and field dependence of the resistivity and Hall effect for the series FeSe${}_{1-x}$S${}_{x}$ x=0.04, 0.09 and 0.19. The sample FeSe${}_{0.81}$S${}_{0.19}$ shows no transition to an orthorhombic phase at low temperatures. Comparison of the transport properties of this sample with the transport properties of other compositions allowed us to reveal the highly mobile electron component that occurred only in an orthorhombic phase which presumably originates from the local FS regions and probably exists in many other orthorhombic phases of iron superconductors.
%	
%%%%%%%%%%%%%%%%%%%%%%%%%%%%%%%%%%%%%%%%%%%%
\section{Experiment}
The crystals of FeSe${}_{1-x}$S${}_{x}$ with x=0.04 and 0.09 were prepared using the KCl/AlCl${}_{3}$ flux technique  \cite{13_CrystEngComm12.1989}. For the sample with x=0.19 the modified method was used described elsewhere \cite{Chareev2016,Chareev2016.2}. The chemical composition of the crystals was studied with the energy dispersive micro analysis system. The composition measurements were done at three points for four average size crystals from the each growth batch. The statistical error in sulfur content was about 5\% for all three batches. 

Electrical measurements were done on cleaved rectangular samples with lengths in the range of 0.5 -- 2 mm, widths about 0.5 mm and thicknesses in the range of 0.01 -- 0.05 mm. The crystal dimensions were measured using a Zeiss binocular microscope. Used method yields up to 30 -- 50\% of systematic errors in  absolute values of resistivity. This error is mainly due to a thickness determination and it does not increase during the transformation from resistivity to conductivity components.
 
Electric contacts were made by sputtering of Au/Ti layers with a precisely machined mechanical mask. The current electrodes were 0.1 mm wide lines along small sides of the bar. Potential and Hall 0.1$\times$0.1 mm$^{2}$ electrodes were connected to a sample holder with a 0.025 mm gold wire using H20E silver epoxy.   The distance between Hall contacts was 0.5 mm.  The distances between current electrodes were 1.5 mm for the sample with x= 0.9 and 1 mm for the samples with x=0.04 and 0.19. Therefore, the corresponding ratios of length to width were 3.0 and 2.0 which give  0.98 and 0.93 for the correction factor due to the Hall potential shortening by extended current electrodes  \cite{15_jan1957galvamomagnetic}. We did not apply these corrections to our data considering it to be insignificant.

DC magnetoresistance and Hall effect measurements were done using QD PPMS and EDX options of MPMS 7T with Keithley 2400 and Keithley 2192. Resistance measurements in pulsed magnetic fields were done up to 25 -- 30 T using 100 kHz AC modulation. The pulse rise time was about 8 msec. 
%
%%%%%%%%%%%%%%%%%%%%%%%%%%%%%%%%%%%%%%%%%%%%%%
\section{Results}
\begin{figure}[h]
\includegraphics[scale=0.5,angle=0]{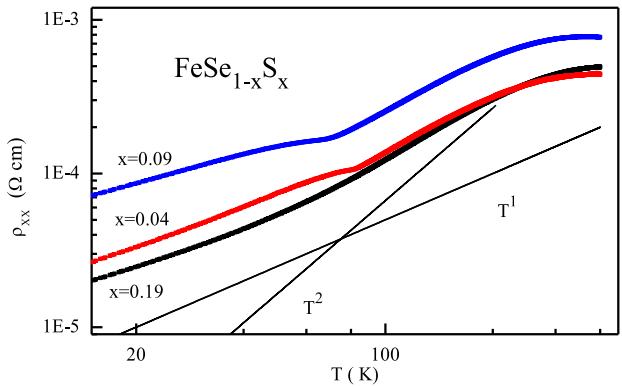}
\caption{\label{fig1} (Color online). Temperature dependence of $\rho_{xx}$ for FeSe${}_{1-x}$S${}_{x}$ (x=0.04, 0.09, 0.19) single crystals.Two lines show the linear and quadratic functions of temperature.  }
\end{figure}
The studied samples show the superconducting transitions at temperatures 10.25, 10.45 and 8.45 K for x=0.04, 0.09 and 0.19 correspondingly. A detailed description of the superconducting properties for these compositions was published elsewhere \cite{16_JLTP}, scanning tunneling microscopy and spectroscopy measurements for x=0.04, 0.09  were reported in Ref.  \cite{PhysRevB.92.235113}. Here we discuss normal state transport properties of these samples. Fig.~\ref{fig1} shows a log-log plot of resistivity versus temperature for the studied FeSe${}_{1-x}$S${}_{x}$ (x=0.04, 0.09, 0.19) samples. Two lines in the plot show slopes of the linear and quadratic power dependencies for the sake of comparison. Discussed above a large systematic error in the sample thickness measurements could offset the curves from the actual positions in the plot.

The shapes of curves reflect qualitative changes in the substance properties under substitution. First of all, the curve corresponding to x=0.19 does not have a kink below 100 K which occurs at the transition to an orthorhombic structure. The absence of the transition in this sample is in agreement with the decrease of the corresponding transition temperature observed for other two samples from 82 K for x=0.04 to 69 K for x=0.09. The similar suppression of structural transition in Fe(SeS) series was also reported by other authors \cite{17_PhysRevB.92.121108}. 

Another peculiarity of $R(T)$ curves is a crossover from a low temperature metal behavior to a saturation at room temperatures. Since the pure FeSe shows a similar saturation that turns into $R(T)$ decrease above 350 K, it can reflect the importance of activated carriers in transport properties of FeSe${}_{1-x}$S${}_{x}$ at high temperatures. Consequently, Gorkov's-Tetelbaum model that describe carriers activation in copper HTSC\cite{18_PhysRevB.77.180511} can be applied to this series of iron superconductors. As it follows from the curves crossing in Fig.~\ref{fig1}, an isovalent substitution of Se by S moves the inflection point of $R(T)$ curve to higher temperatures which may be due to either the increase of the corresponding activation energy or the decrease in the band population.
 
Both the high temperature metal to semiconductor crossover and the low temperature kink due to a structural transition decrease the value of the residual resistance ratio (RRR) determined for $R(T)$ dependencies. Therefore, RRR  underestimates the quality of  the crystals. The resistance of FeSe${}_{0.81}$S${}_{0.19}$ decreases from 300 to 10 K near 15 times reflecting a good quality of the crystal. Furthermore, following the idea of this ratio it is important to note that all $R(T)$ curves plotted  in Fig.~\ref{fig1} are well parallel to each other almost the whole range. It ensures that presented below analysis of the electronic properties of FeSe${}_{1-x}$S${}_{x}$ series was done for the crystals of the same quality.

Our analysis was based on fitting the experimental data with the two or three band model.  To extract parameters it is convenient first to calculate conductivity components from the measured resistivity components. For tetragonal crystals :
\begin{eqnarray}
\sigma_{xx}=\sigma_{yy}=\frac{\rho_{xx}}{(\rho_{xx}^{2}+\rho_{xy}^{2})} \nonumber\\
\sigma_{xy}=\sigma_{yx}=\frac{\rho_{xy}}{(\rho_{xx}^{2}+\rho_{xy}^{2})}
\nonumber %\label{eq:one}
\end{eqnarray}

where $\sigma_{ij}$ are conductivity tensor components and $\rho_{ij}$ are resistivity tensor components. As we described before, the Hall and potential electrodes of our samples were done with precise mask and the main systematic error in measured resistivity components are due to the sample cross-section measurements which equally rescale $\rho_{xy}$ and $\rho_{xx}$ component. Besides, measurements were done in a persistent mode at every $B$, with a current commutation to cancel effect of thermo, contact and other EMF. Mixing of components due to misalignment of potential electrodes and current direction was compensated by extracting of odd and even in magnetic field components for $\rho_{xy}$  and $\rho_{xx}$ correspondingly.  Therefore, we do not expect any additional distortion or error multiplication in the analyzed data due to this transformation.

The conductivities of bands are additive and within relaxation-time approximation for an arbitrary number of bands we can write
\begin{eqnarray}
\sigma_{xx}=F_{R}(B)\equiv\sum_{i=1}^{l}\frac{\lvert\sigma_{i}\rvert}{(1+\mu_{i}^{2}B^2{})}\nonumber\\
\sigma_{xy}=F_{H}(B)\equiv\sum_{i=1}^{l}\frac{\sigma_{i}\mu_{i}B}{(1+\mu_{i}^{2}B^2{})}\nonumber\\
\sigma_{i}=en_{i}\mu_{i}\nonumber
\end{eqnarray}

where  $i$ is a band index, $e$ is the charge of the electron, $\sigma_{i}$ is a conductivity at $B=0$, $\mu_{i}$ is a mobility, $n_{i}$ is a currier concentration and $l$ is a number of bands. Fitting procedure determines $\mu_{i}$ and $n_{i}$ by minimizing the sum:
\begin{equation}
\sum_{k=1}^{N}{\bigg[\bigg({\frac{\sigma_{xx}[k]-  F_{R}(B[k])}{\sigma_{xx}[k]}\bigg)}^{2}+\bigg({\frac{\sigma_{xy}[k]- F_{H}(B[k])}{\sigma_{xy}[k]}\bigg)}^{2} \bigg]} \nonumber
\end{equation}
where $\sigma_{xx}[k]$, $\sigma_{xy}[k]$ and $B[k]$ are the values of  $\sigma_{xx}$, $\sigma_{xy}$ and $B$ at experimental point $k$ and $N$ is the number of the measured points. For the purposes of consistency, we used described fitting method for all samples and for both the two and the three band fitting.  We limited our fitting to temperatures lower or equal to 100 K because at high temperatures the field dependence of $\rho_{xx}$ is too weak to be resolved in the used magnetic field range. For the same reason we used compensated (equal carrier concentration for the hole and electron bands) two band model at 100 K for all samples.
%************************************************************
\subsection{FeSe${}_{0.81}$S${}_{0.19}$}
\begin{figure}[ht]
\includegraphics[scale=0.5,angle=0]{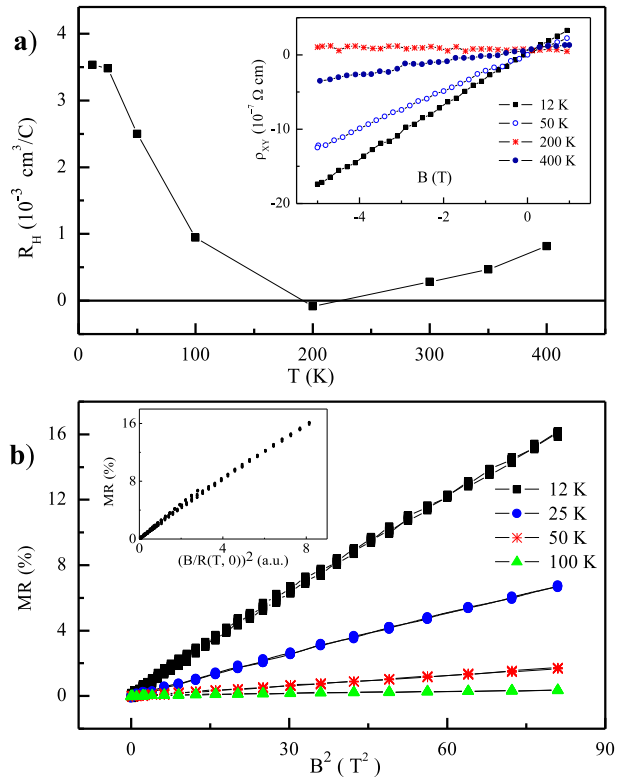}
\caption{\label{fig2}(Color online). (a) Hall coefficient as a function of temperature for  FeSe${}_{0.81}$S${}_{0.19}$. Inset: Magnetic-field dependence of  $\rho_{xy}$ at selected temperatures. (b) The transverse magnetoresistance MR=$(\rho_{xx}(B)-\rho_{xx}(0))/\rho_{xx}(0)$ ( $B \parallel \boldsymbol{c}$) plotted versus $B^{2}$ at temperatures between 12 - 100 K. Inset: MR versus $(B/\rho_{xx}(0))^{2}$ ( Kohler plot) for the same data.  }
\end{figure}
The longitudinal and Hall resistances in a low temperature tetragonal phase behave distinctly different from corresponding resistances of orthorhombic samples. It is in good agreement with the recent report on a quantum critical point near x=0.17 in FeSe${}_{1-x}$S${}_{x}$ series \cite{Hosoi19072016}.  First off all, the orthorhombic FeSe${}_{1-x}$S${}_{x}$ (x=0.04, 0.09) samples show a strongly nonlinear dependence of $\rho_{xy}(B)$ at low temperatures. The $\rho_{xx}(B)$ dependencies also have features which lead, in particular, to a violation of the Kohler's rule, as it will be discussed below. In contrast, the tetragonal at low temperatures composition FeSe${}_{0.81}$S${}_{0.19}$ is closer in properties to a simple compensated semimetal. In support of this assertion, the inset of Fig.~\ref{fig2} (a) shows $\rho_{xy}(B)$ in fields up to 5 T for selected temperatures. All curves are linear which allows to determine the Hall constants at every temperature. The Hall constant shows a strong temperature dependence which is plotted in Fig.~\ref{fig2} (a).  The plotted in Fig.~\ref{fig2} (b) $\rho_{xx}(B)$ dependencies follow $B^{2}$ law and the Kohler's rule is passably satisfied as seen from the inset of Fig.~\ref{fig2} (b).

Our analysis has shown that the measured transport properties of  FeSe${}_{0.81}$S${}_{0.19}$ can be comprehensively described in the framework of the simple two band model. This method allows extracting two band concentrations and mobilities without any additional restrictions on their values. We limited our fitting to temperatures lower or equal to 100 K because of at higher temperatures the field dependence of $\rho_{xx}$ is too weak to be resolved in the used magnetic field range.

\begin{figure}[ht]
\includegraphics[scale=0.5,angle=0]{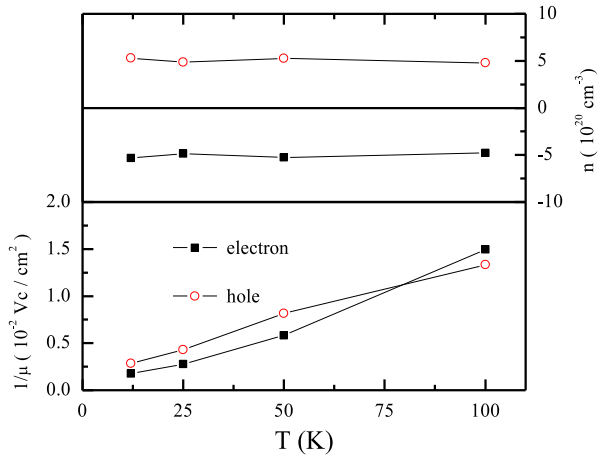}
\caption{\label{fig3}(Color online). The temperature depandence of the carrier density (top panel) and the inverse mobility (bottom panel) for holes and electrones in  FeSe${}_{0.81}$S${}_{0.19}$ extracted from the simple two band model.  }
\end{figure}

The Fig.~\ref{fig3} shows a temperature dependence of carrier concentrations (top) and inverse mobilities (bottom) extracted for FeSe${}_{0.81}$S${}_{0.19}$. The numerical values are listed in Table~\ref{Table1}. The concentration for electron and hole bands are near temperature independent and close to each other. The inverse mobilities show a good linearity without signs of a low temperature saturation that confirms the high quality of the crystal.  
%
%************************************************************
\subsection{FeSe${}_{0.96}$S${}_{0.04}$}
\begin{figure}[h]
\includegraphics[scale=0.5,angle=0]{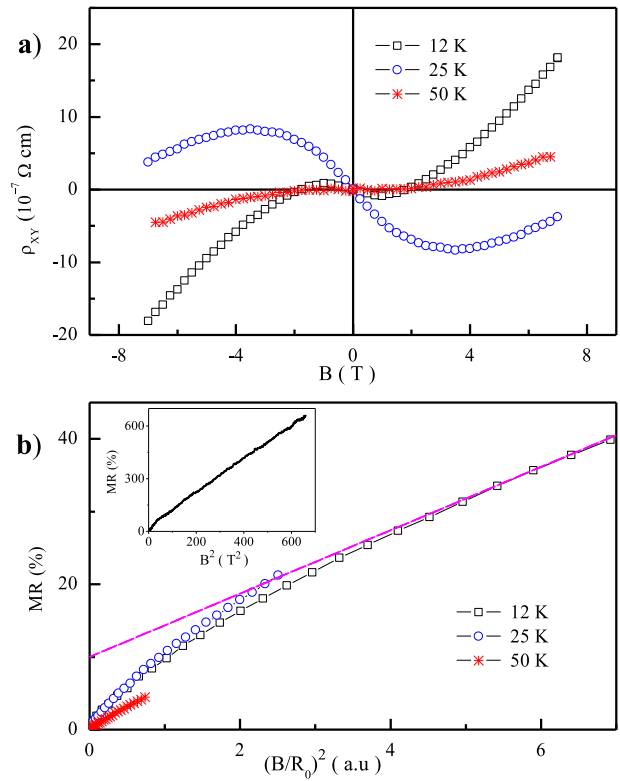}
\caption{\label{fig4}(Color online). (a) Magnetic-field dependence of  $\rho_{xy}$ for FeSe${}_{0.96}$S${}_{0.04}$ at temperatures between 12 - 50 K. (b)  MR versus $(B/\rho_{xx}(0))^{2}$ ( Kohler plot)  for FeSe${}_{0.96}$S${}_{0.04}$ at temperatures between 12 - 50 K and magnetic-field up to 7 T. The straight line highlights the plot curvature. Inset: MR versus $B^{2}$ at 12 K measured in pulsed fields up to 25 T.  }
\end{figure}
The Fig.~\ref{fig4} shows the low temperature $\rho_{xy}(B)$ dependencies (a) and the Kohler's plot (b) under $B$ up to 7 T for the sample with x=0.04. The inset of Fig.~\ref{fig4}(b) shows the high field $\rho_{xx}(B^{2})$ curve measured up to 25 T at 12 K.  The Kohler's rule is not satisfied as it follows from the mismatch of the curve measured at 50 K and curves measured at 12 and 25 K. This conclusion in agreement with the recent study \cite{PhysRevB.93.180503} of Kohler's rule violations for undoped FeSe under pressure. The nonlinear behavior of $\rho_{xy}(B)$ is a clear evidence for the presence of the carriers with a mobility satisfying $\mu_{i}B \approx 1$ in the measured field range.  The $\rho_{xx}(B)$ curve measured at 12 K deviates significantly from $B^{2}$ law below 3 -- 4 T while it follows $B^{2}$ law in higher fields up to 25 T where MR reaches 600 \%  as it shown in the inset of Fig. ~\ref{fig4} (b). It gives reasonable ground to assume existence of a highly mobile band along with the typical for a simple semimetal compensated main hole and electron bands with a considerably lower mobility. The violation of the Kohler's rule in this case may indicate that the properties of the observed highly mobile carriers is a temperature dependent. The conductivity of the highly mobile band is evidently suppressed in relatively low magnetic fields. From Fig. ~\ref{fig4} (b) we can roughly estimate the suppressed part of conductivity near 5 -- 10 \% by the relative amplitude of the low field hump in $\rho_{xx}(B)$ at 12 K. Therefore if we suppose a factor 10 for the mobilities ratio than the density of states for the highly mobile band is only near 1 \% of the overall carrier density.

\begin{figure}[h]
\includegraphics[scale=0.5,angle=0]{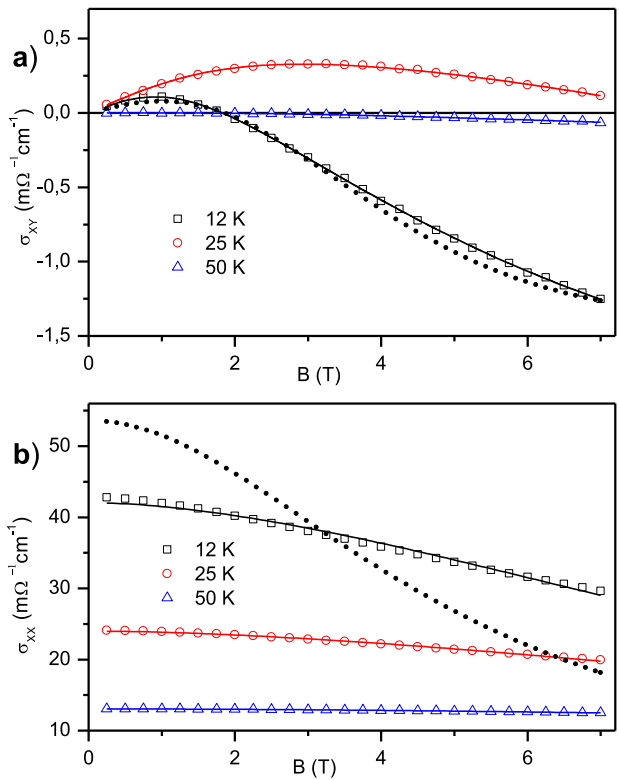}
\caption{\label{fig5}(Color online). The simultaneous fit of $\sigma_{xy}$ (a) and $\sigma_{xx}$ (b) for FeSe${}_{0.96}$S${}_{0.04}$ using the three band model (solid lines). Open circles are experimental data. The dotted lines give the best two band model fit for 12 K data. ($\mu_{n}$=2100 cm$^{2}$/Vs, $\mu_{p}$=1900  cm$^{2}$/Vs, $n_{n}$=7.7 10$^{19}$ cm$^{-3}$, $n_{p}$=9.1 10$^{19}$ cm$^{-3}$) }
\end{figure}

To extract band parameters we fitted the experimental data for FeSe${}_{0.96}$S${}_{0.04}$ with the three band model. All data obtained are listed in Table~\ref{Table1}. The experimental and simulated curves for conductivities at 12, 25 and 50 K are plotted in Fig.~\ref{fig5}. The fit quality is good for all temperatures.  The results confirms the assumption concerning low number of the highly mobile carriers. The obtained value of the carrier density for the highly mobile electrons is near 1$\times$10$^{18}$ cm$^{-3}$ which is two order lower than for the main bands.
%
%************************************************************
\subsection{FeSe${}_{0.91}$S${}_{0.09}$}
\begin{figure}[h]
\includegraphics[scale=0.5,angle=0]{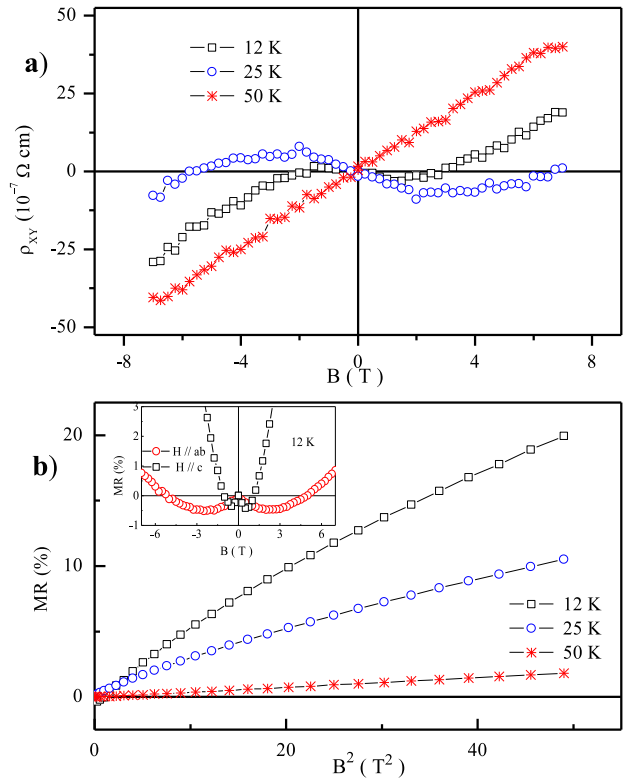}
\caption{\label{fig6}(Color online). (a) Magnetic-field dependence of  $\rho_{xy}$ for FeSe${}_{0.91}$S${}_{0.09}$ at temperatures between 12 - 50 K. (b) MR versus $B^{2}$ for FeSe${}_{0.91}$S${}_{0.09}$. Inset: MR versus B at 12 K for  $B \parallel \boldsymbol{c}$ and $B \parallel \boldsymbol{ab}$}
\end{figure}
%************
The Fig.~\ref{fig6} (a) and (b) show correspondingly $\rho_{xy}(B)$ and $\rho_{xx}(B^{2}$) for the sample with x=0.09. For this sample the value of MR at 12 K in 7 T is near 20 \% which is an intermediate value between 45 \% observed for the sample with x=0.04 and 10 \% for the sample with x=0.19 that confirms systematic changes in electron properties of Fe(SeS) series with sulfur substitution. The low temperature curves are noticeably nonlinear that, similar to the sample with x=0.04, shows the presence of highly mobile carriers. A distinctive feature of the sample with x=0.09 is a negative MR observed at low temperatures. It is likely an occasional property of the particular crystal but it is possibly not the exception.

The inset of Fig.~\ref{fig6} (b) shows dependence of the MR on a magnetic field for the two field orientations. It is clear seen an isotropic behavior of the MR that means that the observed negative MR originates to a scattering suppression. It is usually a magnetic impurity scattering but for FeSe${}_{0.91}$S${}_{0.09}$ the negative MR occurs only in a very narrow temperature range near the onset of superconducting transition (not shown). Based on the last property we suppose that the observed negative MR can originate to the suppression of scattering by superconducting order parameter fluctuations.

The data extracting procedure for this sample was the same as for the sample with x=0.04. Despite the data for FeSe${}_{0.91}$S${}_{0.09}$ is more noisy than for other sample due to the crystal dimensional factor, the fit quality was good for all temperatures. The extracted parameters are listed in Table~\ref{Table1}. Qualitatively, the extracted parameters for the sample FeSe${}_{0.91}$S${}_{0.09}$ are in a good agreement with the parameters for FeSe${}_{0.96}$S${}_{0.04}$.

%********************
\begin{table*}[h]
\caption{\label{Table1}  The results of data fitting using the simple two band model for
 the sample with x=0.19, three band model for samples with x=0.09 and 0.04 at 12, 25 and 50 K and compensated two band model ($n_{h1}=n_{e1}$ ) for samples with x=0.09 and 0.04 at 100 K  }
\begin{ruledtabular}
\begin{tabular}{cccccccc}
 & & \multicolumn{2}{c}{e1}&\multicolumn{2}{c}{h1}&\multicolumn{2}{c}{e2}\\
$x$ & $T $ & $\mu_{n}$ & $n_{n}$ & $\mu_{p}$ & $n_{p}$ & $\mu_{n}$ & $n_{n}$ \\
 &  (K) & (cm$^{2}$/Vs) & (10$^{19}$ cm${}^{-3}$) & (cm$^{2}$/Vs) & (10$^{19}$ cm$^{-3}$) & (cm$^{2}$/Vs) & (10$^{19}$ cm$^{-3}$) \\
 \hline
0.19&12&571&53.4&352&52.9 \\
&25&360&48.7&233&48.6 \\
&50&172&52.6&123&52.6 \\
&100&67&48.0&75&48.0 \\
0.09&12&670&7.0&630&8.0&2100&0.10 \\
&25&410&7.0&430&7.0&1600&0.10 \\
&50&170&10.0&190&12.0&800&0.09 \\
&100&44&27.0&46&27.0 \\
0.04&12&950&12.0&905&15.0&4775&0.12 \\
&25&675&10.0&575&13.0&3398&0.10 \\
&50&265&15.0&298&14.0&1563&0.05 \\
&100&56&38.0&60&38.0 \\
\end{tabular}
\end{ruledtabular}
\end{table*}

%%%%%%%%%%%%%%%%%%%%%%%%%%%%%%%%%%%%%%%%%%%%%%
\section{Discussion and conclusion}
The presence of electrons with considerably higher mobility than those of holes is well known phenomenon for iron-based superconductors. It is usually manifest itself as a low temperature anomaly in field dependence of a resistivity.  For example, BaFe${}_{2}$As${}_{2}$   family demonstrate anomalous low temperature field dependence for both a longitudinal and Hall resistance \cite{7_PhysRevB.84.184514}. Extracted values of  mobility at 5 K were 4500 and 1500 cm$^{2}$/Vs for electron bands and 1800 cm$^{2}$/Vs for a hole band. Another example is a recent paper on FeSe \cite{20_PhysRevLett.115.027006}, where transport property was measured up to 88 T. Authors reports 1843 and 457 cm$^{2}$/Vs for electrons and 623 cm$^{2}$/Vs for holes at 10 K.

Some authors consider a large difference in mobility as a property relating to whole FS pockets.  For example, the study of the Hall effect and resistivity in BaFe${}_{2}$As${}_{2}$ family \cite{21_PhysRevLett.103.057001, 22_PhysRevB.80.140508} showed that electron carriers dominate the transport properties. On this basis the disparity of the electron and hole scattering rates were concluded as universal properties of iron superconductors. Our results for Fe(SeS) series are completely refute this assumption. First of all, the low temperature $\rho_{xy}(B)$ and $\rho_{xx}(B)$ dependencies for the superconducting tetragonal sample FeSe${}_{0.81}$S${}_{0.19}$ do not have anomaly in low magnetic fields. Moreover, this sample behave as a simple two band semimetal. It means that the highly mobile electron band occurs exclusively in an orthorhombic phase. Further, the value of the carrier concentration for the highly mobile electron band  is very low in comparison with other bands. It rather corresponds to local ''hot spots'' than to whole FS pockets.

Our results for the main electron and hole bands of orthorhombic samples are close to the reported for FeSe in Ref. \cite{20_PhysRevLett.115.027006} values while for the highly mobile band our value of mobility is significantly higher and, consequently, carrier concentration is very different. We suspect that the value 1843 cm$^{2}$/Vs was an underestimation due to pure quality of measured crystals. Reported $\rho_{xy}(B)$ curves are almost coincide at 20 and 30 K \cite{20_PhysRevLett.115.027006} which for our mind can be explained only by a high residual resistance of the sample. For our orthorhombic samples  the extremums of $\rho_{xy}(B)$ curves are distinctly moves to low fields with temperature decrease from 25 to 12 K (see Fig.~\ref{fig4}(a) and Fig.~\ref{fig6}(a) ) in agreement with a mobility increase.

Our results for orthorhombic samples are in a good quantitative and qualitative agreement with the results obtained for a high quality FeSe single crystal in Ref. \cite{26_PhysRevB.90.144516} by a mobility spectrum analysis . In the cited paper authors interpret a minority
band with ultrafast carrier mobility as originating either from the Dirac cone or the large anisotropy of FS's. We argue that the large anisotropy of carriers near some local points of FS can be explained by the approach of the Van Hove singularity point to the Fermi level  as a result of a tetragonal to orthorhombic transition.
   
The FS reconstruction at a tetragonal to orthorhombic transition have been intensively studied by both theoretical and experimental methods since the discovery of the iron superconductors. The calculations for FeSe show that in a generic electronic structure consisting of hole pockets/cylinders in the middle of the Brulluen zone (BZ) and electron pockets/cylinders at the border of BZ, the main changes occur at the border of BZ and only minor in the center of BZ \cite{23_PhysRevB.91.214503}. The lifting of degeneracy of $d_{xz}/d_{yz}$ orbitals changes the electron-like cylinders at the border of BZ from round to elliptical form and thus causes the FS shrinking in some directions. We expect that due to a corrugation the cylinder necks can became very thin or even collapse. In both cases the Van Hove singularity which is near the center of the electron cylinders will approach the Fermi level energy. The radius of the electron cylinders necks which can be used as a rough estimate for the distance between the Fermi level and the Van Hove singularity point where determined experimentally. The recent Shubnikov-de Haas oscillation measurements \cite{PhysRevB.90.144517} for FeSe gave the values of the electron neck diameter in the range of  a few meV that means that the Van Hove singularity is in a conduction band even at low temperatures.  The existence of the Van Hove singularity at the Fermi level in FeSe was also discussed in Ref.\cite{24_PhysRevLett.113.237001, 25_PhysRevLett.115.106402}.

Another interesting peculiarity of obtained results is an evolution of the carrier density of the main bands at a tetragonal to an orthorhombic transition. Our analysis (Table~\ref{Table1}),  in agreement with other analyses in the framework of a semiclassical transport theory \cite{20_PhysRevLett.115.027006, 26_PhysRevB.90.144516}, shows the decrease in the electron and hole concentrations at the transition point by several times. It may reflect the overall instability of electronic structure and can be directly related to the bandwidth reductions observed  in iron-based superconductors series \cite{PhysRevX.4.031041, PhysRevLett.115.256403}.

In conclusion, our results demonstrate that a tetragonal to an orthorhombic transition cause an emergence of the highly mobile electrons originating to the local regions of the Fermi surface.     

%%%%%%%%%%%%%%%%%%%%%%%%%%%%%%%%%%%%%%
\begin{acknowledgments}
This work was supported in part from the Ministry of Education and Science of the Russian Federation in the framework of Increase Competitiveness Program of NUST 'MISiS' (№ К2-2015-075 and № K4-2015-020) and by Act 211 of the Government of Russian Federation, agreement № 02.A03.21.0006. We acknowledge support from Russian Foundation for Basic Research Grants № 14-02-00111, 14-02-00245, 14-02-92693, 15-03-99628A, 15-52-45037, ofi-m-16-29-03266, 16-02-00021 and 16-03-00463.
\end{acknowledgments}
\bibliographystyle{apsrev4-1}
\bibliography{FeSeS-mobility_rev.bib}
\end{document}